\documentclass[aps,pre,superscriptaddress,twocolumn,longbibliography]{revtex4}

\usepackage{graphicx,epstopdf,url}
\usepackage[colorlinks=true,urlcolor=blue,citecolor=blue,linkcolor=blue,urlcolor=blue]{hyperref}

\usepackage[active]{srcltx}

\begin{document}
\title{
Friction and mobility of carbon nanoparticles on a graphene sheet
}

\author{Alexander V. Savin}
\email[]{asavin@center.chph.ras.ru}
\affiliation{Semenov Institute of Chemical Physics, Russian Academy of Sciences,
Moscow 119991, Russia}
\affiliation{
Plekhanov Russian University of Economics, Moscow 117997, Russia
}

\begin{abstract}
It is shown using the method of molecular dynamics that the motion of carbon nanoparticles
(rectangular graphene flakes, spherical fullerenes of size $L<10$~nm) on the surface of a thermalized
graphene sheet lying on a flat substrate can be described as the motion of particles in a viscous
medium with a constant coefficient of friction, the value of which depends on the temperature
and particle size.
It has been shown that there are two types of effective friction: diffusion and ballistic.
In ballistic regime of motion (at velocities $v>100$~m/s), deceleration occurs due to the interaction
of moving nanoparticles with thermal out-of-plane bending vibrations of a graphene sheet.
Because of this, with the increasing temperature, the coefficient of friction monotonically increases.
In the diffusion regime of motion (at $v<10$~m/s), friction arises due to the need for the particle
to overcome local energy barriers, therefore it decreases with increasing temperature.
The difference between ballistic and diffusion friction is most pronounced at low temperatures,
since the mobility of nanoparticles in the ballistic regime of motion decreases with increasing
temperature, while in the diffusion regime it monotonously increases.
It is shown that the presence of a normal force pressing the nanoparticle to the substrate
leads to an increases in its friction with the substrate.
\end{abstract}

\maketitle

\section{Introduction}
\label{s1}
Two-dimensional layered materials such as graphene (G, hexagonal boron nitride (hBN),
molybdenum disulfide (MoS$_2$) and tungsten disulfide (WS$_2$) are of great interest because of
their unique electronic \cite{Novoselov2004,Neto2009,Koren2016}
and mechanical \cite{Meyer2007,Lee2008,Falin2017,Han2020} properties.
Due to the very low friction of the layers, these materials can be used as highly efficient dry lubricants
\cite{Sheehan1996,Dienwiebel2004,Lee2010,Cahangirov2012,Liu2012,Yang2013,Koren2015}.
Recently, increased attention has been paid to heterogeneous layered materials that can exhibit
various new physical properties compared to their homogeneous analogues \cite{Leven2013,Geim2013,Novoselov2016}.
Thus, it was shown that the use of G/h-BN heterostructures makes it possible to obtain
the necessary electronic properties \cite{Woods2014,Slotman2015}, as well as significantly
reduce the friction between the layers \cite{Mandelli2017}.

An important task for nano- and micro-sized mechanical devices is to reduce friction
as much as possible \cite{Williams2006}.
Standard lubrication schemes stop working at such sizes, so it is necessary here to switch from 
liquid to dry lubricant, associated with the sliding of flat molecular layers.
This approach, first theoretically proposed several decades ago \cite{Shinjo1993}, allowed to
achieve extremely low coefficients of friction \cite{Hod2018,Martin2018,Baykara2018}.
The use of two-dimensional materials, such as graphene and hexagonal boron nitride h-BN \cite{Berman2018},
makes it possible to achieve extremely low friction.
Layered structures made of these materials can have super-slip layers
\cite{Leven2013,Mandelli2017,Song2018}.

The use of 2D layered structures requires a fundamental understanding of the mechanisms
of the appearance of the friction force at the atomic level.
To date, this mechanism has been well studied only for slow "diffusion" motion of layers
(at velocities $v<10$~m/s) \cite{Vanossi2013,Mandelli2017,Song2018,Ouyang2018,Mandelli2019}.
Modeling of the motion of nanoparticles (clusters Au$_{459}$ \cite{Guerra10}, fullerene molecules C$_{60}$
\cite{Jafary12}) on the surface of a graphene sheet showed that at velocities $v>100$~m/s there is
another "ballistic"\ regime of friction.

The goal of this paper is to explain the mechanisms of the occurrence of friction forces
in the ballistic regime of nanoparticles motion.
For this purpose, the motion of carbon nanoparticles (rectangular graphene flakes and spherical
fullerene molecules) along a thermalized graphene nanoribbon lying on a flat substrate will be simulated.
It will be shown that at high velocities of motion, friction has a "wave"\ origin.
Here, the deceleration of motion occurs due to the interaction of the moving nanoparticle with
thermal out-of-plane bending vibrations of a graphene sheet (the greater are the vibrations,
the higher is the friction).
Therefore, in ballistic motion, unlike standard friction scenarios \cite{Vanossi2013},
the coefficient of friction monotonically increases with increasing temperature.

The paper is structured as follows: the section \ref{s2} describes the full-atomic model,
which is further used to simulate the motion of carbon nanoparticles on a thermalized graphene
sheet lying on a flat substrate.
In section \ref{s3}, the deceleration of the free motion of nanoparticles as a result
of their interaction with the sheet are simulated.
The coefficient of effective friction is determined and the mechanism of its occurrence is analyzed.
In section \ref{s4}, the motion of nanoparticles under the action of a constant force directed
parallel to the substrate surface is modeled.
The mobility of nanoparticles in diffusion and ballistic regimes of motion is analyzed.
In section \ref{s5}, the empirical Amonton-Coulomb law is verified for nanoparticles.
The influence of the normal load (of the force pressing the nanoparticle to the substrate)
on the value of friction is simulated.
The analysis of the results and concluding remarks are given in section \ref{s6}.

\section{Model}
\label{s2}
\begin{figure}[tb]
\begin{center}
\includegraphics[angle=0, width=1\linewidth]{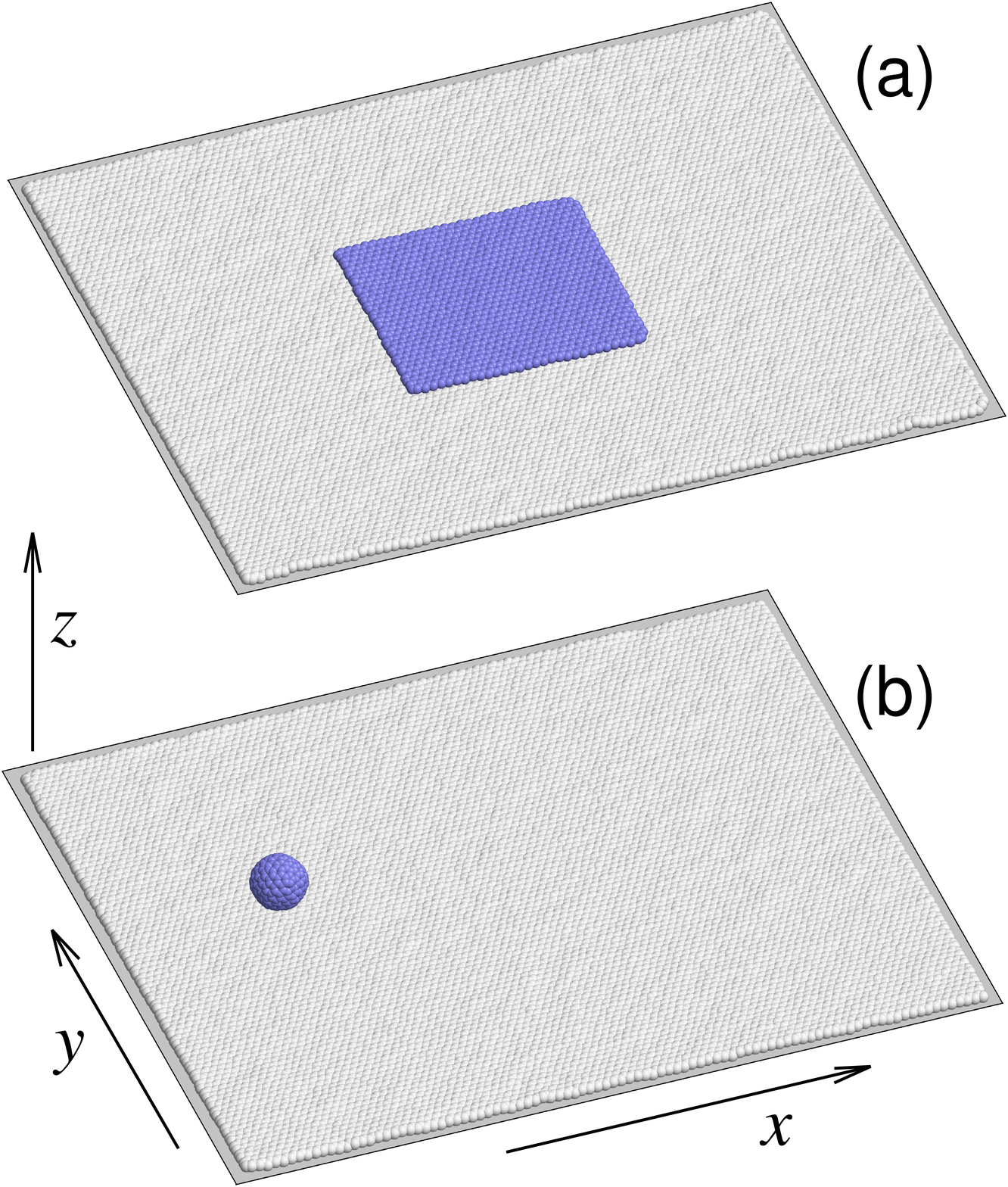}
\end{center}
\caption{\label{fg01}\protect
Simulation of the motion of (a) a rectangular graphene flake (RGF)
of size $7.245\times 6.665$~nm$^2$, consisting of $N_2=1918$ carbon atoms,
and (b) spherical fullerene (SF) C$_{240}$ ($N_2=240$, diameter 1.366 nm),
along a graphene nanoribbon of width 19.001~nm located along the $x$ axis in the $xy$ plane
(the zig-zag direction of the nanoribbon coincides with the $x$ axis, the armchair direction
coincides with the $y$ axis).
The gray color marks the surface of the flat substrate at $z=0$ on which the graphene nanoribbon lies.
}
\end{figure}

Let us simulate the movement of rectangular graphene flakes (RGF) and spherical fullerenes (SF)
along a graphene sheet lying on a flat substrate -- see Fig.~\ref{fg01}.
As a graphene sheet, we will use graphene nanoribbon (GNR) of size 103.0$\times$19.0~nm$^2$
lying in the $xy$ plane (the $x$ axis is directed along the zigzag direction) 
and consisting of $N_1=75598$ carbon atoms.
In realistic cases, the edges of the graphene nanoribbon and rectangular flakes are always
chemically modified.
For simplicity, we assume that the hydrogen atoms are attached to each edge carbon atom.
In our numerical simulations, we take this into account by a change of the mass of the edge atoms.
We assume that the edge carbon atoms have the mass $M_1=13m_p$, while all other internal carbon
atoms have the mass $M_0=12m_p$, where $m_p=1.6601\times 10^{-27}$~kg is the proton mass.

\begin{figure}[t]
\includegraphics[angle=0, width=1\linewidth]{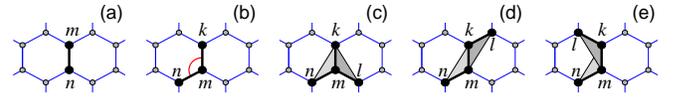}
\caption{(Color online)
Configurations containing up to $i$th type of nearestneighbor interactions for
(a) $i=1$, (b) $i=2$, (c) $i=3$, (d) $i=4$, and (e) $i=5$.
}
\label{fg02}
\end{figure}

Hamiltonian of the RGF(SF)/GNR molecular system  consisting  of $N=N_1+N_2$
carbon atoms ($N_2$ -- number of atoms of RGF (SF)) can be presented in the form,
\begin{equation}
H\!=\sum_{n=1}^{N}\big[\frac12M_{n}(\dot{\bf u}_{n},\dot{\bf u}_{n})+P_{n}+W({\bf u}_n)\big]
+\sum_{n=1}^{N_1}\sum_{k=N_1+1}^{N}\!\!\!\! V(r_{nk}),
\label{f1}
\end{equation}
where $n$ is the number of carbon atom, $M_n$ is the mass of the $n$th atom
(for internal atoms $M_n=12m_p$ and $M_n=13m_p$ for the edge atoms),
${\bf u}_n=(x_n(t),y_n(t),z_n(t))$ is the three-dimensional vector describing the position
of $n$th atom at the time $t$, distance $r_{nk}=|{\bf u}_n-{\bf u}_k|$.
The term $P_n$ describes the interaction of the carbon atom with the index $n$
with the neighboring atoms. The potential depends on variations in bond length, bond angles,
and dihedral angles between the planes formed by three neighboring carbon atoms
and it can be written in the form
\begin{equation}
P=\sum_{\Omega_1}U_1+\sum_{\Omega_2}U_2+\sum_{\Omega_3}U_3+\sum_{\Omega_4}U_4+\sum_{\Omega_5}U_5,
\label{f2}
\end{equation}
where $\Omega_i$, with $i=1$, 2, 3, 4, 5, are the sets of configurations including all interactions
of neighbors. These sets only need to contain configurations of the atoms shown in Fig.~\ref{fg02},
including their rotated and mirrored versions.

Potential $U_1({\bf u}_n,{\bf u}_m)$ describes the deformation energy due to a direct interaction
between pairs of atoms with the indexes $n$ and $m$, as shown in Fig.~\ref{fg02}(a).
The potential $U_2({\bf u}_n,{\bf u}_m,{\bf u}_k)$
describes the deformation energy of the angle between the valence bonds
${\bf u}_n{\bf u}_m$, and ${\bf u}_m{\bf u}_k$, see Fig.~\ref{fg02}(b).
Potentials $U_i({\bf u}_n,{\bf u}_m,{\bf u}_k,{\bf u}_l)$,
$i=3$, 4, and 5, describe the deformation energy associated with a change in the angle between the
planes ${\bf u}_n{\bf u}_m{\bf u}_k$ and ${\bf u}_m{\bf u}_n{\bf u}_k$,
as shown in Figs.~\ref{fg02}(c)--(e).

We use the potentials employed in the modeling of the dynamics of large polymer macromolecules
\cite{Noid1991,Sumpter94} for the valence bond coupling,
\begin{equation}
U_1({\bf u}_1,{\bf u}_2)\!=\!\epsilon_1
\{\exp[-\alpha_0(\rho-\rho_0)]-1\}^2,~\rho\!=\!|{\bf u}_2-{\bf u}_1|,
\label{f3}
\end{equation}
where $\epsilon_1=4.9632$~eV is the energy of the valence bond and $\rho_0=1.418$~\AA~
is the equilibrium length of the bond;
the potential of the valence angle
\begin{eqnarray}
U_2({\bf u}_1,{\bf u}_2,{\bf u}_3)=\epsilon_2(\cos\varphi-\cos\varphi_0)^2,~~
\label{f4}\\
\cos\varphi=({\bf u}_3-{\bf u}_2,{\bf u}_1-{\bf u}_2)/
(|{\bf u}_3-{\bf u}_2|\cdot |{\bf u}_2-{\bf u}_1|),~~
\nonumber
\end{eqnarray}
so that the equilibrium value of the angle is defined as $\cos\varphi_0=\cos(2\pi/3)=-1/2$;
the potential of the torsion angle
\begin{eqnarray}
\label{f5}
U_i({\bf u}_1,{\bf u}_2,{\bf u}_3,{\bf u}_4)=\epsilon_i(1+z_i\cos\phi),\\
\cos\phi=({\bf v}_1,{\bf v}_2)/(|{\bf v}_1|\cdot |{\bf v}_2|),\nonumber \\
{\bf v}_1=({\bf u}_2-{\bf u}_1)\times ({\bf u}_3-{\bf u}_2), \nonumber \\
{\bf v}_2=({\bf u}_3-{\bf u}_2)\times ({\bf u}_3-{\bf u}_4), \nonumber
\end{eqnarray}
where the sign $z_i=1$ for the indices $i=3,4$ (equilibrium value of the torsional angle $\phi_0=\pi$)
and $z_i=-1$ for the index $i=5$ ($\phi_0=0$).

The specific values of the parameters are $\alpha_0=1.7889$~\AA$^{-1}$,
$\epsilon_2=1.3143$ eV, and $\epsilon_3=0.499$ eV, they are found from the frequency
spectrum of small-amplitude oscillations of a sheet of graphite~\cite{Savin08}.
According to previous study~\cite{Gunlycke08}, the energy $\epsilon_4$ is close to
the energy $\epsilon_3$, whereas  $\epsilon_5\ll \epsilon_4$
($|\epsilon_5/\epsilon_4|<1/20$). Therefore, in what follows we use the values
$\epsilon_4=\epsilon_3=0.499$ eV and assume $\epsilon_5=0$, the latter means that we
omit the last term in the sum (\ref{f2}).
More detailed discussion and motivation of our choice of the interaction potentials
(\ref{f3}), (\ref{f4}), (\ref{f5}) can be found in earlier publication~\cite{Savin10}.

The van der Waals interactions of the carbon atoms of the  graphene sheet, flake and fulerene macromolecule
with flat substrate are described by the Lennard-Jones (LJ) potential $(m,l)$
\begin{equation}
W(z)=\epsilon_z[m(z_0/z)^l-n(z_0/z)^m]/(l-m), \label{f6}
\end{equation}
where $z$ is the distance from carbon atom to the outer surface of the substrate,
which is the plane $z=0$. Potential $W(z)$ in Eq. (\ref{f6}) is the interaction energy of a carbon atom
as a function of the distance to the substrate. This energy was found numerically for different substrates
\cite{Savin19,Savin21}. The calculations showed that interaction energy with substrate $W(z)$
can be described with a high accuracy by LJ potential (\ref{f6}) with power $l>k$. Potential (\ref{f6})
has the minimum $W(z_0)=-\epsilon_z$ ($\epsilon_z$ is the binding energy of the atom with substrate).
For the surface of the $\alpha$-graphite crystal, $\epsilon_z=0.052$~eV, $z_0=3.37$~\AA, $l=10$, $m=3.75$.

Nonvalence interactions of the carbon atoms of the nanoribbon and rectangular flake
(fullerene macromolecule) are described by the (6,12) LJ potential
\begin{equation}
V(r)=\epsilon_c\{[(r_c/r)^6-1]^2-1\}, \label{f7}
\end{equation}
where $\epsilon_c=0.002757$~eV, $r_c=3.807$ \cite{Setton96}.

\section{Deceleration of nanoparticles in the ballistic regime of motion}
\label{s3}
It was shown in the papers \cite{Guerra10,Jafary12}  that nanoparticles (cluster Au$_{459}$,
buckminsterfullerene C$_{60}$) moving at high speeds on the surface of a graphene sheet
have a new regime of "ballistic"\ friction, which differs from the well-studied "diffusion"\
friction arising from the drift of nanoparticles at low speeds.
In these regimes, friction is provided by various types of interaction of nanoparticles with the substrate.
To explain the mechanism of "ballistic"\ friction, we will simulate the free motion of carbon
nanoparticles (of RGF and SF) along the graphene nanoribbon (GNR)
of size 103.0$\times$19.0~nm$^2$ (number of carbon atoms $N_1=75598$) lying on a flat substrate.

Let the substrate on which the nanoribbon lies coincide with the plane $z=0$ (nanoribbon plane
coincides with the plane $z=z_0$) and let the nanoribbon lie along the axis $x$.
We place a carbon nanoparticle (RGF or SF) on the center line at the left edge of the nanoribbon
-- see Fig.~\ref{fg01}.
To obtain a normalized state of the molecular system RGF(SF)/GNR, we will place it in a Langevin thermostat.
To do this, we will fix the coordinates $x,y$ of the carbon atoms from the lower left and
upper right corners of the nanoribbon (we put the velocities 
$\dot{x}_n\equiv 0$, $\dot{y}_n\equiv 0$ for $n=1,N_1$).
In order to avoid nanoparticle motion during the thermalization of the system, we will also fix
these coordinates for two opposite angular atoms of RGF and for one atom of SF.
Next, we numerically integrate the Langevin system of equations
\begin{equation}
M_n\ddot{\bf u}_n=-\frac{\partial H}{\partial{\bf u}_n}-\Gamma M_n\dot{\bf u}_n+\Xi_n,~~
n=1,....,N, \label{f8}
\end{equation}
where damping coefficient  $\Gamma=1/t_r$ (time $t_r=0.4$~ps characterizes the intensity
of energy exchange with the thermostat) and
$\Xi_n=\{\xi_{n,i}\}_{i=1}^3$ is three-dimensional vector of normally distributed random
forces (white noise) normalized by conditions
$$
\langle\xi_{n,i}(t_1)\xi_{m,j}(t_2)\rangle=2M_n\Gamma k_BT\delta_{nm}\delta_{ij}\delta(t_2-t_1)
$$
($T$ is thermostats temperature, $k_B$ is Boltzmann constant).

Let us take the initial conditions which correspond to the ground stationary state
of the molecular system "nanoparticle/nanoribbon"\  and numerically integrate the Langevin system
of equations (\ref{f8}) during the time $t_0=20t_r$.
For this time, the system will come into full equilibrium with the thermostat, and we will have
its thermalized state
$$
\{ {\bf w}_n={\bf u}_n(t_0),~~{\bf v}_n=\dot{\bf u}_n(t_0)\}_{n=1}^N.
$$
\begin{figure}[tb]
\begin{center}
\includegraphics[angle=0, width=1\linewidth]{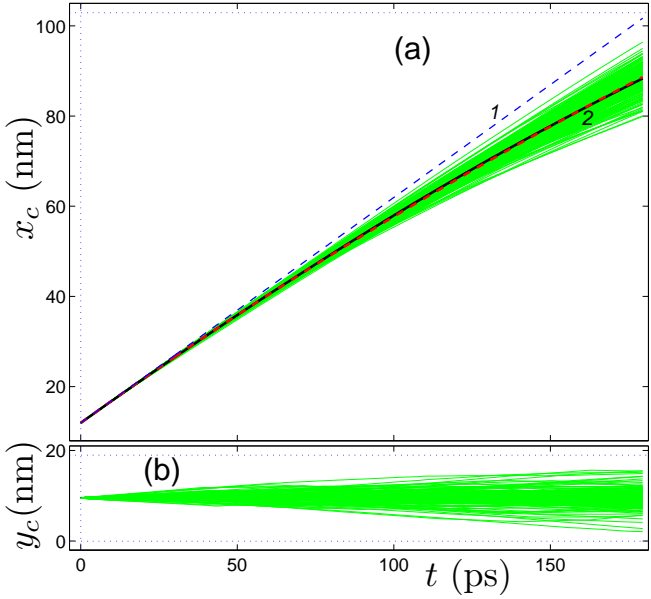}
\end{center}
\caption{\label{fg03}\protect
Trajectories of movement of the center of gravity of RGF of size
3.56$\times$3.26 (number of atoms $N_2=478$) along GNR of size 103$\times$19~nm$^2$
(the number of atoms $N_1=75598$) at a temperature $T=300$K (initial velocity $v_0=500$~m/s).
The green curves show the trajectories of motion for 256 independent realizations of the initial
thermalization of the system, the dashed line 1 is the trajectory for motion with constant
velocity $v=v_0$, the black curve 2 is the averaged trajectory $\bar{x}_c(t)$, and the dashed red
curve is the trajectory of motion (\ref{f12}) with the coefficient of friction $\gamma=1.80$~ns$^{-1}$.
Part (a) shows the dependence of the $x$ coordinate of the center of gravity $x_c$ on time $t$,
part (b) shows the dependence on time for coordinate $y_c$.
Horizontal dotted lines show the boundaries of the nanoribbon.
}
\end{figure}

To simulate the free motion of a nanoparticle along a thermalized infinite graphene nanoribbon,
we leave the interaction with the thermostat only for the nanoribbon atoms located
at a distance less than 1~nm from its left or right edge.
Let us remove all the fixation conditions for the nanoparticle atoms and give them
the additional initial velocity $v_0=500$~m/s directed along the $x$ axis.
To do this, we will numerically integrate the system of equations of motion
\begin{eqnarray}
M_n\ddot{\bf u}_n&=&-\frac{\partial H}{\partial{\bf u}_n}-\Gamma M_n\dot{\bf u}_n+\Xi_n,
\label{f9}\\
&&1\le n\le N_t,~~N_1-N_t<n\le N_1, \nonumber\\
M_n\ddot{\bf u}_n&=&-\frac{\partial H}{\partial{\bf u}_n},
\label{f10}\\
&&N_t<n\le N_1-N_t,~~N_1<n\le N \nonumber
\end{eqnarray}
with initial conditions
\begin{eqnarray}
\label{f11}
{\bf u}_n(0)={\bf w}_n,&&\dot{\bf u}_n(0)={\bf v}_n,~~n=1,...,N_1,\\
{\bf u}_n(0)={\bf w}_n,&&\dot{\bf u}_n(0)={\bf v}_n+v_0{\bf e}_x,~~n=N_1+1,...,N,
\nonumber
\end{eqnarray}
where the number of edge atoms of the nanoribbon interacting with the thermostat is $N_t=900$,
the vector is ${\bf e}_x=(1,0,0)$.

Let us follow the movement of the center of gravity of the nanoparticle along the nanoribbon
$$
x_c=\frac{1}{N_2}\sum_{n>N_1} x_n,~~
y_c=\frac{1}{N_2}\sum_{n>N_1} y_n.
$$
The typical character of the nanoparticle motion is shown in Fig.~\ref{fg03}.
As can be seen in the figure, the type of trajectory $x_c(t)$ depends on the realization
of the initial thermalized state of the system, but interaction with the substrate always
leads to a deceleration of the directional movement of the nanoparticle.
If the trajectory is averaged over all independent realizations of the thermalized state of the system
(if we take the average value $\bar{x}_c(t)=\langle x_c(t)\rangle$), the dynamics can be described
with high accuracy as the motion of a particle in a viscous medium:
\begin{equation}
\bar{x}_c(t)=\bar{x}_c(0)+v_0[1-\exp(-\gamma t)]/\gamma,
\label{f12}
\end{equation}
with the coefficient of viscous "friction"\ $\gamma>0$ (the inverse value $\gamma^{-1}$ corresponds
to the time during which the initial velocity of the nanoparticle decreases $e$ times).
So for RGF of size 3.56$\times$3.26~nm$^2$ at a temperature of $T=300$K, the coefficient of
friction for motion along a graphene sheet $\gamma=1.80$~ns$^{-1}$ -- see Fig.~\ref{fg03}.
In numerical simulation, the averaging of the trajectory was carried out over 256 independent
realizations of the initial thermalized state of the system.
\begin{figure}[tb]
\begin{center}
\includegraphics[angle=0, width=1\linewidth]{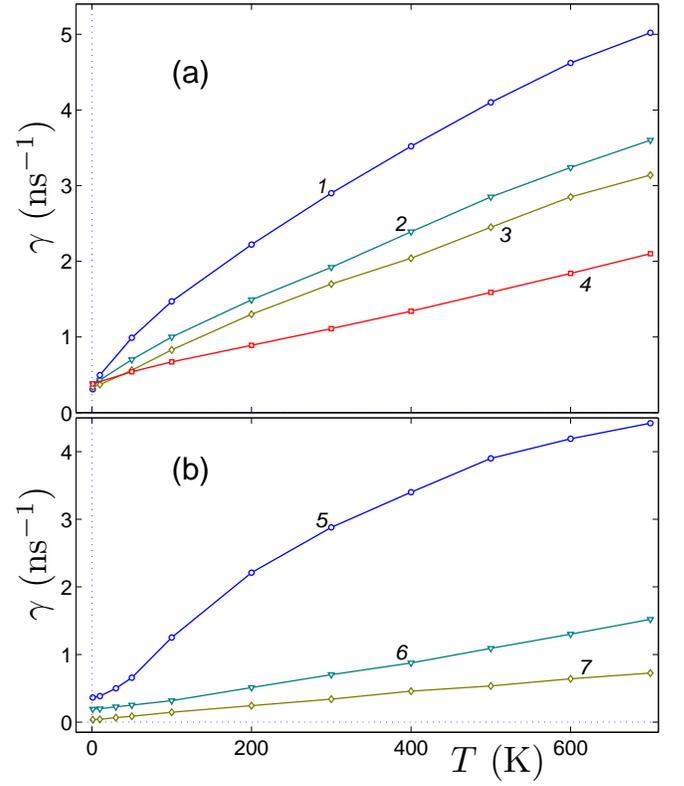}
\end{center}
\caption{\label{fg04}\protect
Dependence of the coefficient of effective friction with substrate $\gamma$ on temperature $T$
for (a) RGF of size: 0.860$\times$0.709, 1.842$\times$1.560,
3.561$\times$3.261 and 7.245$\times$6.665~nm$^2$ (curves 1, 2, 3 and 4) and for (b)
SF C$_{20}$, C$_{60}$ and C$_{240}$ (curves 5, 6 and 7).
}
\end{figure}

Numerical simulation of nanoparticle dynamics has shown that in ballistic regime
viscous deceleration (\ref{f12})
occurs at all nanoparticle sizes and temperatures $T\ge 100$K, but the value of the effective friction
coefficient $\gamma$ depends on the temperature, size and type of nanoparticle -- see Fig.~\ref{fg04}.
The coefficient of friction monotonically increases with the increase in temperature and decreases with
the increase in nanoparticle size.
For large nanoparticles, the coefficient of friction increases linearly with the increase in temperature.
This allows us to conclude that the deceleration of the ballistic motion of
nanoparticles is caused by their interaction with the thermal vibrations of the substrate.
It should be noted that the "diffusion"\ regime of nanoparticle motion is characterized by
a decrease in friction with an increase in temperature, due to additional thermal activation
of their jumps through local energy barriers \cite{Prandtl28}.
\begin{figure}[tb]
\begin{center}
\includegraphics[angle=0, width=1\linewidth]{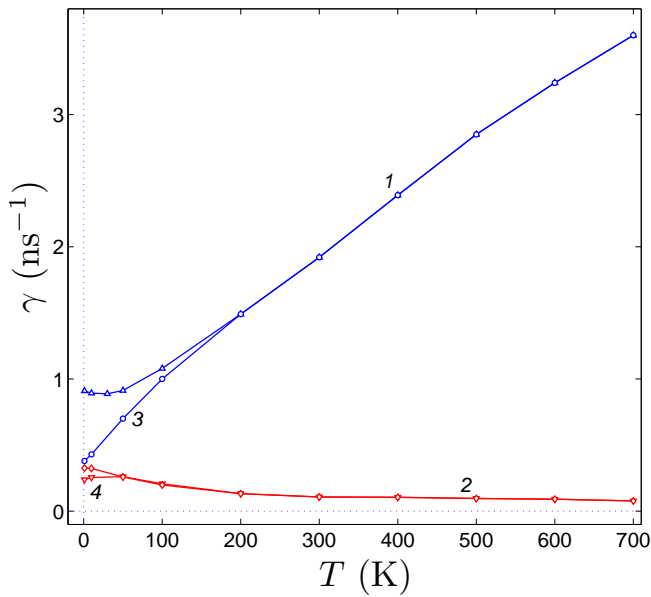}
\end{center}
\caption{\label{fg05}\protect
Dependence of the effective coefficient of friction with the substrate $\gamma$ on temperature $T$
for RGF of size $1.842\times 1.560$~nm$^2$ by taking into account all three-dimensional
movements of atoms (curves 1, 3) and by taking into account only movements in the $xy$ plane
(curves 2, 4).
Curves 1, 2 show dependencies by choosing the direction of GRF movement which is commensurate with 
substrate. Curves 3 and 4 illustrate dependencies by choosing incommensurate direction of GRF movement.
}
\end{figure}

For the diffusion of RGF on a graphene layer, the transitions of the flake from commensurate to incommensurate
states with the layer are very important \cite{Lebedeva10,Liu14}.
Let us estimate the contribution of these transitions to friction for the ballistic regime of RGF motion.
In our numerical simulation, the zigzag graphene nanoribbon model an infinite graphene layer.
Zigzag directions of GNR and RGF coincide with the $x$ axis.
Therefore, by shifting the flake along the $y$ axis, we can obtain both directions
of RGF movement that are commensurate and incommensurate to the substrate.

Numerical simulation of the motion of a graphene flake of size 1.842$\times$1.560~nm$^2$ has shown
that the choice of a commensurate direction leads to an increase in the coefficient
of friction only at low temperatures $T<100$K -- see Fig.~\ref{fg05}.
At higher temperatures, the commensurability of the direction of the flake ballistic movement
to the graphene layer does not affect the value of the coefficient of friction.
Here, an increase in temperature leads to a monotonous, almost linear, increase in friction.

If, when modeling the ballistic movement of a flake, extra-plane displacements of atoms are prohibited,
i.e. if we switch from a three-dimensional to a two-dimensional model, 
the value of friction will decrease by one order of magnitude and the direction of the temperature 
dependence of the friction coefficient will change on opposite.
When using a 2D model for the ballistic regime, as well as for the diffusion regime of motion,
the coefficient of friction will decrease monotonically with the increase in temperature -- see Fig.~\ref{fg05}.
Therefore, it can be concluded that the friction in the ballistic regime of motion has a wave nature.
The reason for the deceleration is the interaction of a moving nanoparticle with thermal
out-of-plane bending vibrations of a graphene sheet.
The greater are these fluctuations, the greater is the value of the coefficient of friction.

Let us note that for a rectangular flake, the initial sliding velocity above the used value 
$v_0=500$~m/s can be obtained as a result of it partially shift beyond the edge of the substrate.
Then the return of the flake onto the substrate will lead to its sliding at velocity $v=600$~m/s \cite{Liu14}.
\begin{figure}[tb]
\begin{center}
\includegraphics[angle=0, width=1\linewidth]{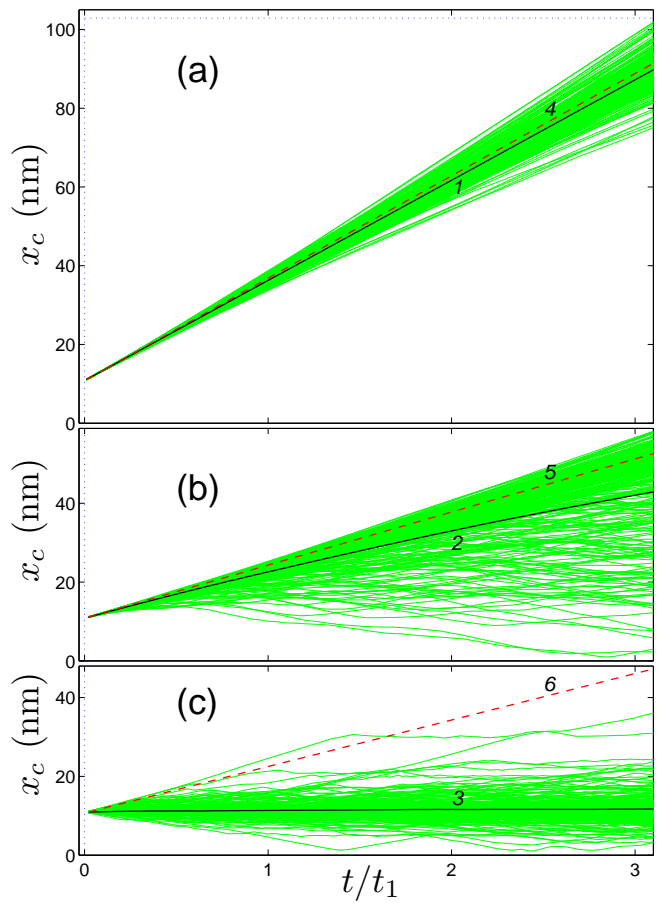}
\end{center}
\caption{\label{fg06}\protect
Trajectories of movement of the mass center of RGF of size 1.842$\times$1.560~nm$^2$
($N_2=126$) under the action of constant force $F=F_c/N_2$ at (a)$F_c=0.00045$, (b) 0.00025
and (c) 0.0001~eV/\AA~ (temperature $T=100$K, time $t_1=100$, 92.9 and 200~ps).
The green curves show the trajectories of motion for 256 independent realizations of the initial
thermalization of the system, the black curves 1, 2, 3 are averaged trajectories $\bar{x}_c(t)$,
dashed lines 4, 5, 6 are trajectories for motion with constant velocity $v_0=N_2F/M_c\gamma$,
where the coefficient of friction for ballistic regime of motion $\gamma=0.00108$~ps$^{-1}$.
}
\end{figure}

\section{Mobility of nanoparticles}
\label{s4}
If a particle of mass $M_c$ moves in a viscous medium along the $x$ axis under the action
of an external force $F_c$, then its dynamics is described by the equation of motion
\begin{equation}
M_c\ddot{x}_c=-\gamma M_c\dot{x}_c+F_c,
\label{f13}
\end{equation}
where $\gamma$ is the coefficient of friction characterizing the viscosity of the medium.
From the equation (\ref{f13}) follows that over time the particle will always enter
the regime of motion with a constant velocity $v(F_c)=F_c/\gamma M_c$.
Here, the mobility of the nanoparticle, the ratio of the constant velocity of its movement to the force,
\begin{equation}
\mu=v(F_c)/F_c=1/\gamma M_c
\label{f14}
\end{equation}
completely determines the coefficient of friction $\gamma$.
\begin{figure}[tb]
\begin{center}
\includegraphics[angle=0, width=1\linewidth]{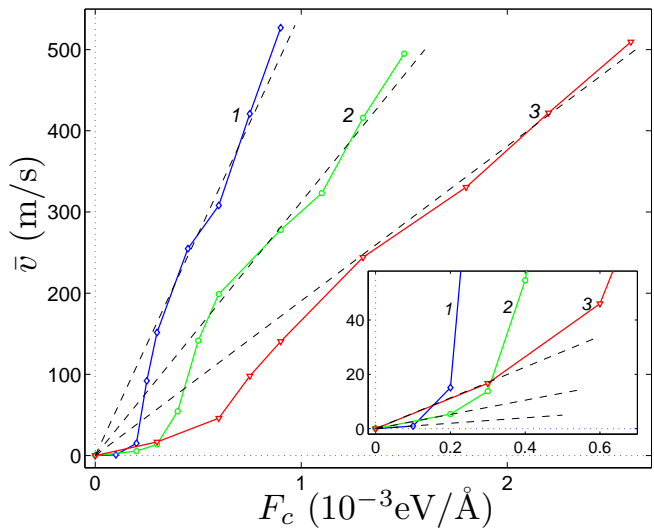}
\end{center}
\caption{\label{fg07}\protect
Dependence of the average value of the velocity of the nanoparticle $\bar{v}$
on the force $F_c$ for RGF of size 1.842$\times$1.560~nm$^2$ at a temperature
$T=100$, 300 and 600K (curves 1, 2 and 3).
The dashed lines show the dependencies corresponding to the mobility of the nanoparticle
in the ballistic and diffusion (in the inset) regimes of motion.
}
\end{figure}
\begin{figure}[tb]
\begin{center}
\includegraphics[angle=0, width=1\linewidth]{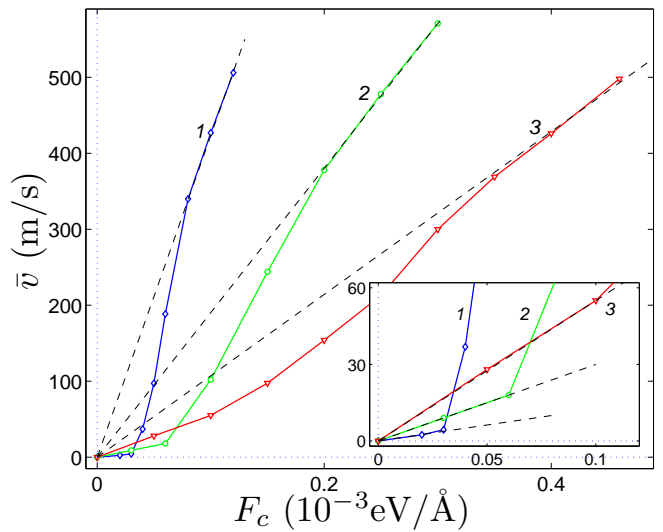}
\end{center}
\caption{\label{fg08}\protect
Dependence of the average value of the velocity of the nanoparticle $\bar{v}$
on the force $F_c$ for SF C$_{60}$ at a temperature
$T=100$, 300 and 600K (curves 1, 2 and 3).
The dashed lines show the dependencies corresponding to the mobility of the nanoparticle
in the ballistic and in the diffusion (in the inset) regimes of motion.
}
\end{figure}

Let us find the mobility of a carbon nanoparticle on a graphene sheet $\mu$ through direct modeling
of its motion on thermalized sheet under the action of a constant force $F>0$ applied to each
atom of the nanoparticle parallel to the $x$ axis.
In this case, the dynamics of the particle will be described by a system of equations of motion
\begin{equation}
M_n\ddot{\bf u}_n=-\frac{\partial H}{\partial {\bf u}_n}+F{\bf e}_x,~~N_1<n\le N.
\label{f15}
\end{equation}

The total mass of the particle is $M_c=\sum_{n=1}^{N_2} M_{N_1+n}$,
the total force acting on its center of gravity is $F_c=N_2F$,
so we should expect that the velocity of a uniform motion of the particle $v_0(F)=N_2F/M_c\gamma$.
Thus, to simulate the forced motion of a nanoparticle, it is necessary to numerically integrate
the system of equations of motion (\ref{f9}), (\ref{f10}), (\ref{f15}) with an initial condition
(\ref{f11}), where the velocity of the initial motion $v_0=N_2F/M_c\gamma$.

In the simulation, the trajectory of the particle's center of gravity was averaged
over 256 independent realizations of the initial thermalized state of the system
$\{ {\bf w}_n,{\bf v}_n\}_{n=1}^N$, and the value of the constant velocity
of movement $\bar{v}$ was determined from this trajectory.
A form of the trajectories of the nanoparticle motion under the action
of a constant force is shown in Fig.~\ref{fg06}.
Numerical simulation has shown that the nanoparticle has two values of mobility $\mu$.
At high values of the external force, the particle always enters the ballistic regime
of motion with constant velocity $\bar{v}\approx v_0(F)$ -- see Fig.~\ref{fg06}~(a).
Here the mobility is $\mu\approx 1/M_c\gamma$, where $\gamma$ is the coefficient
of friction for ballistic regime of motion.
At small values of force, the particle always enters the diffusion regime of motion
with average velocity value $\bar{v}\ll v_0(F)$ -- see Fig.~\ref{fg06}~(c).
At intermediate values of force, both diffusive and ballistic motion of the particle
is possible -- see Fig.~\ref{fg06}~(b).

The ballistic mobility of a nanoparticle always significantly exceeds the diffusion mobility.
For example, for RGF of size 1.842$\times$1.560~nm$^2$ (for fullerene C$_{60}$) at a temperature
$T=100$K, mobility for the ballistic regime exceeds that for the diffusion regime of motion 
by 60 (35) times, at $T=300$K by 12 (5.2) times, and at $T=600$K by 3.5 (1.9) times.

The form of the dependence of the average value of the nanoparticle velocity $\bar{v}$
on the value of the constant force $F_c$ is shown in Fig.~\ref{fg07} and \ref{fg08}.
The velocity value always monotonically increases with the increase in force.
For high and low values of force, the velocity of motion becomes directly proportional to the force:
for small values $\bar{v}\sim\mu_dF_c$, and for large values $\bar{v}\sim\mu_bF_c$
with proportionality coefficients $\mu_d\ll\mu_b$ ($\mu_d$ and $\mu_b$ is the mobility
of the particle in diffusion and ballistic regimes of motion).
Thus, the simulation shows that in both regimes the particle motion is realized in a viscous medium,
but with different coefficients of friction.
In the diffusion mode, the effective friction caused by the interaction of the nanoparticle
with the substrate is significantly higher than the friction in the ballistic mode:
$\gamma_d=1/M_c\mu_d\gg\gamma_b=1/M_c\mu_b$.

With the increase in temperature, friction decreases monotonically in the diffusion regime,
and it increases in the ballistic regime  of motion, which indicates different mechanisms
of its occurrence.
Therefore, the difference between the diffusion and ballistic regimes of motion is most
pronounced at low temperatures.
The simulation shows that at speed of motion $v<10$~m/s the nanoparticle always enters
the diffusion regime and at $v>100$~m/s -- the ballistic regime of motion.
\begin{figure}[tb]
\begin{center}
\includegraphics[angle=0, width=0.96\linewidth]{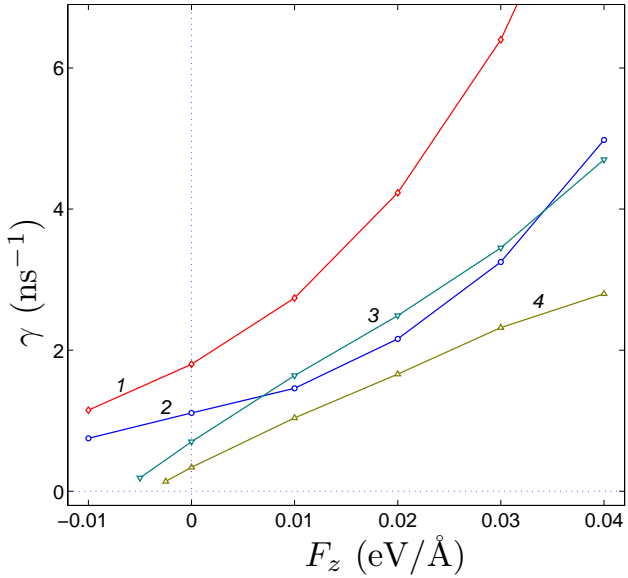}
\end{center}
\caption{\label{fg09}\protect
Dependence of the coefficient of effective friction with the substrate $\gamma$
on the force of the normal load $F_z$ for ballistic motion of RGF of size:
3.561$\times$3.261, 7.245$\times$6.665~nm$^2$ (curves 1, 2) and of SF
C$_{60}$, C$_{240}$ (curves 3, 4). Temperature $T=300$K.
}
\end{figure}
\begin{figure}[tb]
\begin{center}
\includegraphics[angle=0, width=1\linewidth]{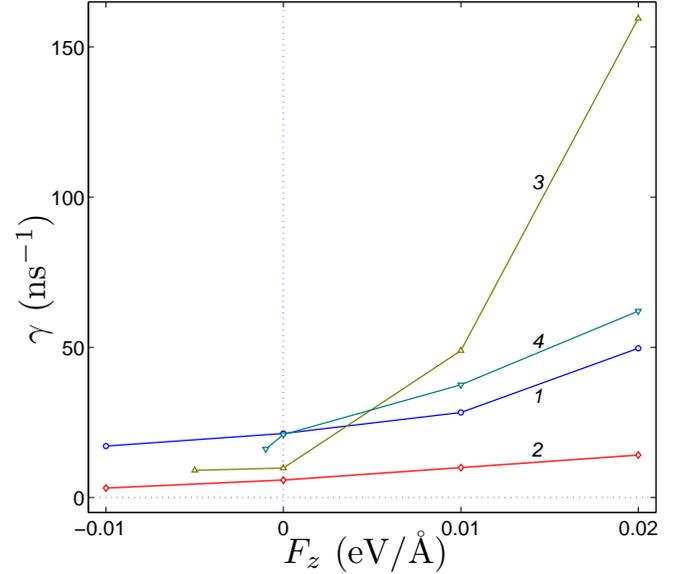}
\end{center}
\caption{\label{fg10}\protect
Dependence of the coefficient of effective friction with the substrate $\gamma$
on the force of the normal load $F_z$ for diffusive motion of RGF of size:
1.84$\times$1.56, 3.561$\times$3.261 (curves 1, 2) and for SF C$_{60}$, C$_{240}$ (curves 3, 4).
Temperature $T=300$K.
}
\end{figure}

\section{Effect of normal load on friction}
\label{s5}
According to the empirical Amonton-Coulomb law, the friction force increases with the increase
in the normal load.
Recent papers  \cite{Deng12,Smolyanitsky12,Sun18,Mandelli2019} have shown that for
layered structures such as graphite, this law may not be fulfilled at the nanoscale.
An increase in the normal load can lead to a decrease in the friction force between the layers.

Let us check the Amonton-Coulomb law for nanoparticles located on a graphene sheet.
For this, we will simulate their movement taking into account a force $F_z$ that presses it vertically
to the substrate.
In this case, the dynamics of the nanoparticle will be described by a system of equations of motion
\begin{equation}
M_n\ddot{\bf u}_n=-\frac{\partial H}{\partial {\bf u}_n}+F_z{\bf e}_z,~~N_1<n\le N,
\label{f16}
\end{equation}
where vector ${\bf e}_z=(0,0,1)$.
Thus, to simulate the ballistic motion of a nanoparticle taking into account a normal load $F_z$,
it is necessary to numerically integrate a system of equations of motion
(\ref{f9}), (\ref{f10}), (\ref{f16}) with an initial condition (\ref{f11}), where the velocity $v_0=500$~m/s.

Numerical simulation of dynamics has shown that the addition of a normal force $F_z$ pressing
the nanoparticle to the substrate leads to a monotonous increase in the coefficient of friction
$\gamma$ in both ballistic and diffusion regimes of motion -- see Fig.~\ref{fg09}, \ref{fg10}.
The rate of growth of the coefficient of friction decreases with the increase in particle size.
Thus, for the nanoparticles considered in this paper, the Amonton-Coulomb law is fulfilled.

It is not yet possible to verify the validity of the law for large nanoparticles
using a full-atom model due to computational difficulties associated with very large dimensions
of systems of equations of motion.
However this can be done if, instead of a 3D model, we use a 2D model that requires 
significantly less computing resources.
It has been shown by using this model \cite{Savin21a} that by ballistic motion of large graphene flakes,
the friction force acts differently on the edge and on the inner atoms of the flake.
For the edges of the flake, friction always increases with an increase in the normal load
due to the indentation of the edges into the substrate, and for the inner part, friction decreases
due to additional alignment of the surfaces of the flake and the substrate.
The inner sections will make the main contribution to friction only for flakes of size $L>250$~nm,
therefore, only for flakes of this large size should we expect a decrease in friction
with an increase in normal load.

\section{Conclusion}
\label{s6}
The numerical simulation of motion of carbon nanoparticles (rectangular graphene flakes, spherical fullerenes)
on the surface of a graphene sheet lying on a flat substrate has shown that there are two regimes
of friction: diffusion and ballistic.
In these regimes, effective friction takes place due to different types of interaction of the
nanoparticle with a graphene sheet.
In the well-studied diffusion regime (for velocities $v<10$~m/s), friction occurs due
to overcoming local energy barriers.
Here, the value of friction depends on the commensurability of the direction of motion
of the nanoparticle with the lattice of the sheet, and the friction coefficient always
decreases monotonically with the increase in temperature.
In the ballistic regime (for $v>100$~m/s), the commensurability of the direction of motion
does not affect on friction, and the friction coefficient 
increases monotonically, almost linearly, with the  increase in temperature.
Here, the friction is of a wave nature.
The cause of deceleration is the interaction of a moving nanoparticle with thermal
out-of-plane bending vibrations of a graphene sheet (the greater are these vibrations,
the higher is the friction).

Simulation of the motion of nanoparticles under the action of a constant force has shown
that their mobility in the ballistic regime (at large values of force) is
higher than in the diffusion regime (at small values of force).
With an increase in temperature in the diffusion regime, the mobility monotonically increases,
and it decreases in the ballistic regime.
Therefore, the difference between the diffusion and ballistic regimes of motion is most
pronounced at low temperatures.

The simulation of motion in the presence of a normal force pressing the nanoparticle to the substrate
has shown that the increase in the normal load leads to a monotonous increase in the coefficient
of friction in both ballistic and diffusion regimes of motion.
The rate of growth of the coefficient of friction decreases with increasing particle size,
but it always remains positive.
It can be concluded that for particles with sizes $L<10$~nm located on a graphene sheet,
the empirical Amonton-Coulomb law is always valid (the violation of the law should be expected
for particles of size $L>250$~nm).

\begin{center}
{\bf ACKNOWLEDGMENTS}
\end{center}

Computational facilities were provided by the Interdepartmental Supercomputer Center of the Russian
Academy of Sciences.

\end{document}